\documentclass[12pt,draft]{article}

\usepackage{amsfonts}

\newcommand{\Io}{{\mathbb I}}
\newcommand{\Ro}{{\mathbb R}}

\newcommand{\be}{\begin{eqnarray}}
\newcommand{\ee}{\end{eqnarray}}

\renewcommand{\Re}{{\,\rm Re\,}}
\renewcommand{\Im}{{\,\rm Im\,}}

\newcommand{\kk}{{\bf k}}
\newcommand{\qq}{{\bf q}}

\newcommand{\xvar}{x}
\newcommand{\yvar}{y}

\begin{document}

\title{Model of the quantized electromagnetic field
in the presence of sources}

\author{
Jan Naudts and
Wojciech De Roeck\footnote{Aspirant Fonds voor Wetenschappelijk Onderzoek Vlaanderen}\\
\small Departement Natuurkunde, Universiteit Antwerpen UIA,\\
\small Universiteitsplein 1, 2610 Antwerpen, Belgium;\\
\small E-mail: Jan.Naudts@ua.ac.be and Wojciech.DeRoeck@ua.ac.be.
}

\date{}

\maketitle

\begin{abstract}
We give a rigorous description of a model of
the quantized electromagnetic field interacting with quantized current fields.
In the special case of classical currents our results agree with common
knowledge about the problem. A toy model of a quantum current is
studied as well.

\end{abstract}



\section{Introduction}


The traditional approach to quantum electrodynamics starts with
the theory of free photons, respectively free electrons. Next,
interactions are introduced. They are treated using scattering
theory, making the assumption that
particles are free in the large time limit.
It is well known that the resulting theory suffers from intrinsic
difficulties, like infrared and ultraviolet divergences.
Here we show how to describe in a rigorous manner an electromagnetic (e.m.)
field interacting with quantized current fields. In principle, these
fields might be generated by Dirac electron fields. However, we prefer
to avoid the difficulties of fully quantized electrodynamics
by delaying the rigorous formulation of electron fields
to a forthcoming paper. The integration of
both models into a mathematically acceptable theory of QED remains an open
problem.

Important for the rigorous formulation of the problem at hand
is the choice of method for quantizing the
e.m.~field. As shown quite some time ago \cite {CGH77},
the method of Fermi can be made rigorous. The essence of \cite {CGH77}
has been picked up in \cite {KN02} and was used to give
an elegant formulation of quantized free e.m.~fields
using the formalism of covariance systems \cite {NK01}.
In particular, free-field operators are smeared out using test
functions satisfying the continuity equation. By doing
so they become gauge independent \cite {KTWB68} and the difficult
problem of gauge covariance of the quantized theory
is avoided.

Next a model is needed for the current fields.
Given any such fields one can construct vector potential fields
by integrating the currents with the Green's function of the
d'Alembert equation. These vector potential fields
can be described in a similar way as the free-field potentials.
This is the basis for the present paper.
The quantized e.m.~fields and the currents are described
simultaneously by a single covariance system. An essential property of
such a covariance system is that only minimal information is needed
about the properties of the two subsystems and of their interaction,
because such information is encoded in the states of the system,
which are determined by their correlation function.
To illustrate this point we discuss states describing
quantized e.m.~fields
interacting with classical currents and show that in this case
we recover known results. But we discuss also states
describing genuine interaction with arbitrary current fields.


The structure of the paper is as follows. Sections 2, 3 and 4 serve
as a technical introduction. Our basic {\sl ansatz} is made in
section 5. Section 6 contains two propositions exploring the
consequences of this ansatz. Sections 7 and 8 show that
our results are in agreement with results of the standard approach.
In section 9 we introduce a simplified model of quantum currents.
The physical interpretation of radiation fields of this model
follows in section 10. Finally, section 11 draws
some conclusions. Two appendices contain
some further technical matters.

\section{Test functions}

In photon theory a number of pitfalls have to be avoided and
some choices have to be made.
To begin with, it is well-known that the field operators $\hat A_\mu(q)$
do only exist in a distributional sense. Therefore, smear\-ed-out operators
are defined by
\be
\hat A(f)=\int_{\Ro^4}{\rm d}q\,f^\mu(q)\hat A_\mu(q).
\label{smo}
\ee
The $f_\mu(q)$ are test functions reflecting the experimental
inaccuracy to select a single point of Minkowski space.
The Fourier transformed function $\tilde f_\mu(k)$
satisfies
\be
f_\mu(q)=(2\pi)^{-2}\int_{\Ro^4}{\rm d}k \tilde f_\mu(k)e^{-iq^\nu k_\nu}.
\ee
From the reality of $f_\mu(q)$ follows
that $\overline{\tilde f_\mu(k)}=\tilde f_\mu(-k)$.

Following \cite {KTWB68,KN02}, we assume that
the test functions satisfy the Fourier-transformed continuity equation
\be
k^\mu \tilde f_\mu(k)=0.
\label{wfl}
\ee
A justification is given in Appendix A.
This assumption is essential in the present formalism to control
problems with gauge invariance of the theory.

In what follows the space of test functions $f_\mu$ is
denoted $G$. It consists of real functions
$f_\mu(q)$ whose Fourier transform satisfies (\ref{wfl}).
For technical reasons, we assume that
the Fourier transformed functions are continuous and vanish
outside a bounded region.

\section{Classical wave functions}


It is tradition to introduce a complex Hilbert space of
so-called classical wave functions of the photon.
The space $G$ is only a real space, because of the condition
that $f_\mu(q)$ is real, while the classical wave functions
form a complex pre-Hilbert space $H$. These are square integrable complex functions
$\phi_\mu(\kk)$, defined for $\kk$ in $\Ro^3$, satisfying
\be
|\kk|\phi_0(\kk)=\sum_{\alpha=1}^3k_\alpha\phi_\alpha(\kk).
\label{lcsp}
\ee
With each test function $f_\mu(q)$ corresponds a classical wave function
$\phi_\mu(\kk)$ by the relation
\be
\phi_\mu(\kk)=\sqrt{2\pi}\tilde f_\mu(|\kk|,\kk),
\qquad \kk\in\Ro^3.
\ee


The (degenerate) scalar product for classical wave functions is given by
\be
\langle\phi|\psi\rangle
&=&-\int_{\Ro^3}{\rm d}\kk\,
\frac{1}{2|\kk|}
\overline{\phi^\mu(\kk)}\psi_\mu(\kk).
\label {defscalprod}
\ee
Positivity of this scalar product follows because the classical wave functions
satisfy (\ref{lcsp}). Indeed, one has
\be
\langle\phi|\phi\rangle
&=&-\int_{\Ro^3}{\rm d}\kk\,\frac{1}{2|\kk|^3}
\left(|\kk|^2|\phi_0(\kk)|^2-|\kk|^2\sum_{\alpha=1}^3|\phi_\alpha(\kk)|^2\right)\cr
&=&\int_{\Ro^3}{\rm d}\kk\,\frac{1}{2|\kk|^3}
\sum_{\alpha,\beta=1}^3\left[
\delta_{\alpha\beta}|\kk|^2-k_\alpha k_\beta\right]
\overline{\phi_\alpha(k)}\phi_\beta(k)\cr
&\ge&0.
\ee
The latter holds because the matrix with elements
$\delta_{\alpha\beta}|\kk|^2-k_\alpha k_\beta$
is positive definite for all values of $\kk$.
We conclude that the classical wave functions form a
pre-Hilbert space $H$.

\section{Correlation function description of\\ e.m. fields}

The smeared-out field operators, which will be constructed later on,
satisfy the canonical commutation relations
\be
\left[\hat A(f),\hat A(g)\right]_-
=-2i\Im \langle\phi|\,\psi\rangle,
\ee
where $\phi$ and $\psi$ are the classical wave functions
determined by $f$ and $g$, respectively. The displacement
operators are defined by
\be
\hat W(f)=\exp(i\hat A(f)).
\ee
They satisfy the Weyl form of commutation relations
\be
\hat W(f)\hat W(g)=e^{i\Im \langle\phi|\,\psi\rangle}\hat W(f+g)
\label{weyl}
\ee
and generate an algebra which is {\sl not} the algebra of canonical
commutation relations \cite {PD90} because the symplectic form
\be
f,g\rightarrow \Im \langle\phi|\,\psi\rangle
\ee
is clearly degenerate.

We need several distinct representations of this algebra. Free photon fields are described by
field operators in Fock space. On the other hand, if the presence of
an external current produces an infrared divergency, then a representation is needed
which differs from the Fock representation. The obvious way to handle such a situation is
by means of correlation functions. They determine the Hilbert space representation
uniquely by means of the G.N.S.~representation theorem.


Instead of working with mathematical states of the $C^*$-algebra of
canonical commutation relations \cite {PD90} we work with the
correlation function formalism of \cite {NK01}. The basic quantity is the two-point correlation
function ${\cal F}(f,g)$, defined for any pair of test functions $f$ and $g$ in $G$.
In a Hilbert space representation
with state vector $\Omega$ it has the following meaning
\be
{\cal F}(f,g)=\left\langle \hat W(g)^*\Omega|\,\hat W(f)^*\Omega\right\rangle.
\label{repr}
\ee
The scalar product between two vectors $\Phi$ and $\Psi$ of the Hilbert space
is denoted $\langle\Psi|\,\Phi\rangle$ and is complex linear in $\Phi$,
anti-linear in $\Psi$.

From (\ref{repr}) it is obvious that the correlation function ${\cal F}(f,g)$
is nothing but the inner product between two coherent states, one with state vector
$\hat W(f)^*\Omega$, the other with state vector $\hat W(g)^*\Omega$. The characterizing
properties of correlation functions are in the present context
\begin{itemize}
\item (normalization) ${\cal F}(0,0)=1$;
\item (positivity) $\sum_{j,k}\lambda_j\lambda_k{\cal F}(f_j,f_k)\ge 0$ for all finite
sequences of complex numbers $\lambda_1,\cdots,\lambda_n$ and of
elements $f_1,\cdot f_n$ of $G$.
\item (covariance) There exists a symplectic form $\sigma$ over $G$ such that
\be
{\cal F}(f+h,g+h)=e^{i\sigma(f-g,h)}{\cal F}(f,g)
\ee
holds for all $f,g,h$ in $G$.
\end{itemize}

The general definition of states of a covariance system contains also a requirement
of continuity. However, the additive group of test functions $G,+$ is
equipped with the discrete topology. Hence continuity of the map
$f,g\rightarrow {\cal F}(f,g)$ is always satisfied.


In particular, the correlation function for the vacuum state of the free photon field is given by
\cite {KN02}
\be
{\cal F}(f,g)
&=&e^{\langle \phi|\,\psi\rangle}
e^{-(1/2)\langle \phi|\, \phi\rangle}
e^{-(1/2)\langle \psi|\,\psi\rangle}\cr
&=&e^{i\sigma(f,g)}e^{-(1/2)s(f-g,f-g)}
\label{corfun}
\ee
with
\be
\sigma(f,g)&=&\Im\langle \phi|\,\psi\rangle\cr
s(f,g)&=&\Re\langle \phi|\, \psi\rangle.
\ee
Here, $\phi$ and $\psi$ are the classical wave functions determined by
$f$ and $g$, respectively.
Expression  (\ref{corfun}) satisfies all requirements for being a correlation function
--- see Appendix B.
By the generalized GNS-theorem \cite {NK01} there exists a projective representation $\hat W(f)$
of the group $G$ in a Hilbert space $\cal H$, and a vector $\Omega$ in
$\cal H$, such that (\ref{repr}) holds.

\section{Describing quantized currents}


The currents $\hat j_\mu(q)$, given as operator-valued distributions in a
Hilbert space $\cal H$, can be used to construct vector potentials
$\hat A^{\rm j}_\mu(q)$ by
\be
\hat A^{\rm j}_\mu(q)=-\int{\rm d}q'\,\Delta_G(q'-q)\hat j_\mu(q'),
\ee
where $\Delta_G(q)$ is a Green's function of the d'Alembert equation,
i.e.~is a solution of
\be
\square_q\Delta_G(q)=-\delta^{4}(q).
\ee
One can take $\Delta_G(q)$ equal to Feynman's propagator for massless bosons
(\ref {fprop}). The formal equation
\be
\square_q\hat A^{\rm j}_\mu(q)=\hat j_\mu(q)
\ee
is satisfied by construction.
After smearing out with test functions one obtains
\be
\hat A^{\rm j}(f)=-\int{\rm d}q\,\int{\rm d}q'\,f^\mu(q)\Delta_G(q'-q)\hat j_\mu(q').
\label {smoutcp}
\ee
Given these operators one can reconstruct the current fields by
applying the d'Alembert operator. Indeed, introduce the notation
\be
\tau_af_\mu(q)=f_\mu(q-a).
\ee
Then one has formally
\be
\square_a \hat A^{\rm j}(\tau_af)
&=&\int{\rm d}q\,f^\mu(q-a)\hat j_\mu(q)\cr
&\equiv&\hat j(\tau_af).
\ee


Note that the free-field operators $\hat A(f)$ satisfy the homogeneous equation
\be
\square_a \hat A(\tau_af)
&=&0.
\ee
As a consequence, two free-field operators $\hat A(f)$ and $\hat A(g)$
are equal if they determine the same classical wave function (modulo
the null-space of $H$). This property is in general {\sl not} true
for the operators $\hat A^{\rm j}(f)$.
Nevertheless one can use the formalism of covariance systems to describe these
$\hat A^{\rm j}(f)$ in a similar way as for describing free field operators.
Introduce the notation
\be
\hat W^{\rm j}(f)=\exp(i\hat A^{\rm j}(f)),
\ee
and assume that a correlation function ${\cal F}^{\rm j}(f,g)$ is given
such that in the G.N.S.-representation one has
\be
{\cal F}^{\rm j}(f,g)=
\left\langle \hat W^{\rm j}(g)^*\Omega|\,\hat W^{\rm j}(f)^*\Omega\right\rangle.
\label{reprj}
\ee
The current operators $\hat j(f)$ are fully specified by this correlation function.
Examples of such functions follow below.

\section{Interacting fields}


Let us now construct an interacting field operator $\hat A^{\rm I}_\mu(q)$
which is the sum of the free field operator $\hat A_\mu(q)$ and of the
field $\hat A^{\rm j}_\mu(q)$ produced by the current. The latter two
are described by the correlation functions ${\cal F}(f,g)$,
respectively ${\cal F}^{\rm j}(f,g)$. These have to be combined into a
single correlation function ${\cal F}^{\rm I}(f,g)$, the G.N.S.-representation
of which contains Weyl operators satisfying
\be
\hat W^{\rm I}(f)=\exp(i\hat A^{\rm I}(f))=\exp(i(\hat A(f)+\hat A^{\rm j}(f))).
\ee


An easy way to produce correlation functions with the desired properties
starts from correlation functions ${\cal F}^\times(f,f';g,g')$
of the covariance system with group $G\times G$. By taking the
diagonal of such a function one obtains a correlation function
of the covariance system with group $G$
\be
{\cal F}^{\rm I}(f,g)={\cal F}^\times(f,f;g,g).
\ee
The G.N.S.-representation induced by ${\cal F}^{\rm I}(f,g)$
can be obtained from that induced by ${\cal F}^\times(f,f;g,g)$.


The simplest class of correlation functions of the product
system consists of functions of the form
\be
{\cal F}^\times(f,f';g,g')
&=&\exp\left(i\sigma^\times(f,f';g,g')\right)\cr
& &\times
\exp\left(-\frac{1}{2}s^\times(f-g,f'-g';f-g,f'-g')\right)
\label {prodcorfun}
\ee
with $s^\times(f,f';g,g')$ a real inner product of $G\times G$
and with $\sigma^\times(f,f';g,g')$ a symplectic form of $G\times G$
such that
\be
(f,f';g,g')=s^\times(f,f';g,g')+i\sigma^\times(f,f';g,g')
\label {posreq}
\ee
defines a positive-definite sesquilinear form of $G\times G$.
These are analogues of the quasi-free states of \cite{PD90}.
Interaction between e.m.~field and current is supposed to be
such that
\be
{\cal F}^\times(f,0;g,0)&=&{\cal F}(f,g),\cr
{\cal F}^\times(0,f';0,g')&=&{\cal F}^{\rm j}(f',g').
\label {freecon}
\ee

Let the G.N.S.-representation induced by ${\cal F}^\times(f,f';g,g')$ satisfy
\be
{\cal F}^\times(f,f';g,g')
=\left\langle \hat W^\times(g,g')^*\Omega|\,\hat W^\times(f,f')^*\Omega\right\rangle.
\ee
Introduce generators $\hat A^{\rm I}(f)$, $\hat A(f)$, and $\hat A^{\rm j}(f)$ by
\be
\hat W^\times(f,f)&=&\hat W^{\rm I}(f)=\exp(i\hat A^{\rm I}(f))\cr
\hat W^\times(f,0)&=&\exp(i\hat A(f))\cr
\hat W^\times(0,f)&=&\exp(i\hat A^{\rm j}(f))
\ee
By construction is $\hat A^{\rm I}(f)=\hat A(f)+\hat A^{\rm j}(f)$.
The commutation relations for the operators $\hat W^{\rm I}(f)$
are
\be
\hat W^{\rm I}(f)\hat W^{\rm I}(g)=e^{i\sigma^{\rm I}(f,g)}\hat W^{\rm I}(f+g).
\ee
The symplectic form appears in the r.h.s.~of the commutation relations
\be
\left[\hat A^{\rm I}(f),\hat A^{\rm I}(g)\right]_-
&=&-2i\sigma^\times(f,f;g,g)=-2i\sigma^{\rm I}(f,g)\cr
\left[\hat A(f),\hat A(g)\right]_-
&=&-2i\sigma^\times(f,0;g,0)\cr
\left[\hat A^{\rm j}(f),\hat A^{\rm j}(g)\right]_-
&=&-2i\sigma^\times(0,f;0,g)\cr
\left[\hat A(f),\hat A^{\rm j}(g)\right]_-
&=&-2i\sigma^\times(f,0;0,g).
\label {comrel}
\ee

\section{Classical currents}


In the simplest case the currents $\hat j_\mu(q)$ are multiples of the
identity operator. Then the operators $\hat W^{\rm j}(f')$ and $\hat W^{\rm j}(g')$
can be taken out of the inner product of (\ref {reprj}). A suitable guess is therefore
\be
{\cal F}^{\rm j}(f,g)=\exp\left(iA^{\rm cl}(g-f)\right)
\label {classcurcor}
\ee
with
\be
\hat A^{\rm cl}(f)=-\int{\rm d}q\,\int{\rm d}q'\,f^\mu(q)\Delta_G(q'-q) j_\mu(q').
\label {ikg}
\ee
It is straightforward to verify that this function satisfies all requirements
for being a correlation function.


Eq.~(\ref{classcurcor}) allows very general classical potentials.
Take e.g.~the Coulomb potential
\be
A_\mu^{\rm cl}(q)=\delta_{\mu,0}\frac{c}{|\qq|},
\label{coulombpot}
\ee
where $c$ is a constant, and where $q=(q_0,\qq)$.
Equations (\ref{ikg}) are satisfied with
external current
\be
j_\mu(q)=\delta_{\mu,0}4\pi c\delta^3(\qq).
\ee
This means that a static charge of strength $4\pi c$ is
located at the origin of space.
The occurrence of a divergency in (\ref{coulombpot})
does not produce any problem because it enters (\ref{classcurcor}) in a form
smeared out with test functions.


Note that $\sigma^{\rm j}(f,g)=0$. The commutation relations (\ref {comrel})
suggest to take
\be
\sigma^\times(f,0;0,g)=\sigma^\times(0,f';0,g)=0.
\ee
It is still possible
to include non-trivial correlations between the classical currents
and the free-field operators by means of the inner product $s^\times(f,f';g,g')$. However,
standard results about quantized e.m.~fields in presence of
classical currents are recovered when the choice
\be
\sigma^\times(f,f';g,g')&=&\sigma(f,g)+A^{\rm cl}(g-f)\cr
s^\times(f,f';g,g')&=&s(f,g).
\ee
is made.

All together, the correlation function of the classical current model reads
\be
{\cal F}^{\rm I}(f,g)
={\cal F}(f,g){\cal F}^{\rm j}(f,g)
\label{intcorfun}
\ee
with ${\cal F}(f,g)$ given by (\ref {corfun})
and ${\cal F}^{\rm j}(f,g)$ given by (\ref {classcurcor}).
Such a simple product form reflects the known fact that the
classical current model does not contain any interactions.
By this is meant that in a Heisenberg picture
the Hamiltonian is the sum of two free parts, without
additional interaction term.


In the present model the interacting field operator
equals the sum of the free-field operator and the classical
potential generated by the external current
\be
\hat A^{\rm I}(f)=\hat A(f)+A^{\rm cl}(f)\hat\Io.
\label{explic}
\ee
This property is known in literature --- see e.g.~Eq.~2.63 of \cite {PS95}.
Still, many handbooks use the classical current model to illustrate the
scattering approach of QED, without mentioning (\ref {explic}).
For sake of completeness we discuss some results of the scattering
context in the next section.


Note that the field operators $\hat A^{\rm I}(f)$ have some unusual
properties. From (\ref{explic}) is clear that they satisfy the same
canonical commutation relations as the free-field operators.
But the representation depends intrinsically
on the details of the external current. Indeed,
a shift in spacetime may map a non-zero field operator
onto zero. This means that the
shifted representation is not unitary equivalent with the
original representation.

Let us analyze this point in somewhat more detail.
A field operator $\hat A^{\rm I}(f)$ vanishes if and only if
$A^{\rm cl}(f)=0$ and $\langle\phi|\,\phi\rangle=0$,
where $\phi$ is the classical wave function associated with $f$.
Now, if the current $j_\mu(q)$ is not trivial, then
there exists a test function $f$ for which
$A^{\rm cl}(f)\not=0$ and $\langle\phi|\,\phi\rangle=0$ holds.
If the current is localized in part of spacetime then
shift the test function with a vector $a$ so that $\tau_a f$
vanishes in that part of spacetime where the current does not vanish.
The result is that $A^{\rm cl}(\tau_a f)=0$. Because
$f$ and $\tau_a f$ determine the same classical wave function
up to a phase factor one concludes that $\hat A^{\rm I}(f)\not=0$
while $\hat A^{\rm I}(\tau_af)=0$.

\section {Radiation fields}


Let us first verify what happens if $A^{\rm cl}(f)$ is a solution of the
homogeneous d'Alembert equation, i.e.~the current vanishes. Then one can
write
\be
A^{\rm cl}(f)
&=&\int_{\Ro^3}{\rm d}\kk\,\frac{1}{2|\kk|}
\left(a^\mu(\kk)\overline{\phi_\mu(\kk)}+\overline{a^\mu(\kk)}\phi_\mu(\kk)
\right)\cr
&=&-2\Re\langle a|\,\phi\rangle,
\label {vccl}
\ee
with $\phi$ the classical wave function determined by the test functions $f$
and with
\be
a_\mu(\kk)=\frac{1}{\sqrt{2\pi}}\tilde A_\mu^{\rm cl}(|\kk|,\kk).
\ee
Hence the correlation function can be written as
\be
{\cal F}^{\rm I}(f,g)
&=&{\cal F}(f,g)\exp\left(-2i
\Re\langle a|\,\psi-\phi\rangle\right)\cr
&=&\exp\left(i\Im\langle \phi+2ia|\,\psi+2ia\rangle\right)
\exp(-(1/2)\langle\phi-\psi|\,\phi-\psi\rangle).\cr
& &
\ee
In this expression the Fourier transformed classical potential $a$,
when multiplied with $2i$, behaves as a Fourier transformed
test function. This relation is a duality between test functions
and fields and has been studied in \cite{KN02}.
In particular, the definition of field operators $\hat A(f)$
can be extended to complex arguments. Hence one can write
\be
{\cal F}^{\rm I}(f,g)
&=&{\cal F}\left(f+i\pi^{-1}A^{\rm cl},g+i\pi^{-1}A^{\rm cl}\right)\cr
&=&\left\langle \hat W(g+i\pi^{-1}A^{\rm cl})^*\Omega|\,
\hat W(f+i\pi^{-1}A^{\rm cl})^*\Omega\right\rangle\cr
&=&\langle\Omega(A^{\rm cl})|\,\hat W(g)\hat W(f)^*\Omega(A^{\rm cl})\rangle,
\ee
with
\be
\Omega(A^{\rm cl})=\hat W(i(2\pi)^{-1}A^{\rm cl})^*\Omega.
\ee
To obtain the latter use that
$\hat W(f+2g)=\hat W(g)\hat W(f)\hat W(g)$.
This shows that in this case the correlation functions (\ref{intcorfun})
are those of a coherent state, as expected from conventional
photon theory.


Next let us make a link with the scattering approach.
Consider a classical field $A_\mu^{\rm cl}(q)$ which vanishes for
very negative times $q_0<<0$. This implies that $j_\mu(q)=0$
for very negative times. Assume that $j_\mu(q)=0$ holds also for very
positive times $q_0>>0$. Then $A_\mu^{\rm cl}(q)$ for $q_0>>0$
is the radiation field produced by currents $j_\mu(q)$ which are
only active during a finite interval of time $q_0$. This radiation
field can be expressed in terms of the current using Feynman's
propagator
\be
D_F(q)
&=&\frac{1}{(2\pi)^4}\int_{\Ro^4}{\rm d}k\,\frac{1}{k^\nu k_\nu}e^{-ik^\mu q_\mu}.
\label {fprop}
\ee
which is a Green's function of the d'Alembert equation.
One finds
\be
A^{\rm cl}_\mu(q)
&=&A^{\rm hom}_\mu(q)
-\int{\rm d}q'\,j_\mu(q')D_F(q-q')\cr
&=&A^{\rm hom}_\mu(q)
-\frac{1}{(2\pi)^2}\int_{\Ro^4}{\rm d}k\,
\frac{1}{k^\nu k_\nu}e^{-ik^\lambda q_\lambda}\tilde j_\mu(k)
\ee
where $A^{\rm hom}_\mu(q)$ is a solution of the homogeneous d'Alembert
equation. Because of the assumption that $A^{\rm cl}_\mu(q)=0$
when $q_0<<0$, one must have
\be
A^{\rm hom}_\mu(q)
&=&
\frac{1}{(2\pi)^2}\int_{\Ro^4}{\rm d}k\,
\frac{1}{k^\nu k_\nu}e^{-ik^\lambda q_\lambda}\tilde j_\mu(k),
\qquad q_0<<0.
\ee
The latter equation can be written as
\be
A^{\rm hom}_\mu(q)
&=&-\frac{1}{2\pi}\Im\int_{\Ro^3}{\rm d}\kk\,\frac{1}{2|\kk|}
e^{i|\kk|q_0}e^{i\kk\cdot\qq}\tilde j_\mu(|\kk|,\kk).
\ee
In this form the expression is valid for all $q$. Indeed, one checks
immediately that this is a solution of the homogeneous d'Alembert
equation. For $q_0<<0$ one finds $A^{\rm cl}_\mu(q)=0$ by
construction, for $q_0>>0$ one has $A^{\rm cl}_\mu(q)=2A^{\rm hom}_\mu(q)$.

Smearing out $A^{\rm hom}_\mu(q)$ with a test function $f$ one obtains
(see (\ref{vccl}))
\be
A^{\rm hom}(f)=\frac{1}{2\pi}\Im\langle \phi|\,a\rangle
\ee
with $\phi$ the classical wave function determined by the test functions $f$,
and with
\be
a_\mu(\kk)=\frac{1}{\sqrt{2\pi}}\tilde j_\mu(|\kk|,\kk).
\ee
Hence, for test functions with support lying far in the future
one has
\be
A^{\rm cl}(f)=\frac{1}{\pi}\Im\langle \phi|\,a\rangle.
\ee
The correlation function (\ref{intcorfun}) for a pair of
such functions reads
\be
{\cal F}^{\rm I}(f,g)
&=&{\cal F}(f,g)\exp\left((i/\pi)\Im\langle \psi-\phi|\,a\rangle\right).
\label{intcorfunscat}
\ee
Compare this with
\be
& &\langle \hat W(g)^*\hat W(v)^*\Omega|\,
\hat W(f)^*\hat W(v)^*\Omega\rangle\cr
&=&\exp(i\Im\langle \phi-\psi|\,\xi\rangle)
\,\langle\hat W(v+g)^*\Omega|\,\hat W(v+f)^*\Omega\rangle\cr
&=&\exp(i\Im\langle \phi-\psi|\,\xi\rangle){\cal F}(v+f,v+g)\cr
&=&\exp(2i\Im\langle \phi-\psi|\,\xi\rangle){\cal F}(f,g),
\ee
where $\xi$ is the classical wave function corresponding with the
test function $v$. The two expressions
coincide provided there exists a test function $v$ such that
$\xi=-a/2\pi$ holds. If this is the case then the radiation field
is described by the coherent state with wave vector $\hat W(v)^*\Omega$.
This coincides with the standard result that, up to a phase factor,
the S-matrix is a displacement operator, and the radiative field is
a coherent state --- see e.g.~\cite {BB75}, section 13.
The problem is not the appearance of a complex test function,
which has been explained above, but the possibility of an infrared
divergency. Indeed, a test function $v$, such that $\xi=-a/2\pi$ holds,
will not always exist. One can of course try to approximate
$a_\mu(\kk)=\tilde j_\mu(|\kk|,\kk)/\sqrt{2\pi}$ by classical
wave functions $\xi_n$. However, this will work only if
\be
\langle a|\,a\rangle
&=&-\frac{1}{2\pi}\int_{\Ro^3}{\rm d}\kk\,\frac{1}{2|\kk|}
\overline {\tilde j^\mu(|\kk|,\kk)}\tilde j_\mu(|\kk|,\kk)
\ee
is finite. This is precisely the condition for
absence of infrared divergency found in \cite {BB75}, section 13, in case
of an example.
For recent progress on the infrared divergency problem in the context
of Nelson's model see \cite{AA01,LMS02}

\section{Quantum currents}


Let us now consider a simplified model of quantum currents.
Start with creation and annihilation operators $\hat b^*$ and $\hat b$
of a harmonic oscillator. They satisfy the canonical commutation relations
\be
\left[\hat b,\hat b^*\right]_-=1
\ee
Let be given complex functions $\alpha_\mu(q)$ satisfying the
continuity equation
\be
\partial_\mu\alpha^\mu(q)=0.
\ee
They are used to define currents $\hat j_\mu(q)$ by the relation
\be
\hat j_\mu(q)=\alpha_\mu(q)\hat b^*+\overline {\alpha_\mu(q)}\hat b.
\ee
The smeared-out potentials (\ref {smoutcp}) become
\be
\hat A^{\rm j}(f)=\yvar(f)\hat b^*+\overline{\yvar(f)}\hat b
\ee
with
\be
\yvar(f)
=-\int{\rm d}q\,\int{\rm d}q'\,f^\mu(q)\Delta_G(q'-q)\alpha_\mu(q').
\label {yform}
\ee
They satisfy commutation relations
\be
\left[\hat A^{\rm j}(f),\hat A^{\rm j}(g)\right]_-
&=&-2i\sigma^{\rm j}(f;g)
\ee
with
\be
\sigma^{\rm j}(f,g)=\Im\left(
\overline {\yvar(f)}\yvar(g)
\right).
\ee


Let $\Omega$ denote the ground state of the harmonic oscillator.
It satisfies $b\Omega=0$ and determines the correlation
function ${\cal F}^{\rm j}(f,g)$ via (\ref {reprj}).
One obtains by means of a
standard calculation
\be
{\cal F}^{\rm j}(f,g)
&=&\exp\left(-i\sigma^{\rm j}(f,g)\right)
\exp\left(
-(1/2)s^{\rm j}(f-g,f-g)\right)
\label {fj}
\ee
with
\be
s^{\rm j}(f,g)=\frac{1}{2}\left(\yvar(f)\overline{\yvar(g)}+\yvar(g)\overline{\yvar(f)}\right).
\ee
In this state the quantum expectation of the current $\hat j_\mu(q)$
vanishes. The second moment equals
\be
\langle\Omega|\,\hat j_\mu(q)\hat j_\nu(q')\Omega\rangle
=\overline{\alpha_\mu(q)}\alpha_\nu(q').
\ee

The interaction between photons and currents is modeled by assuming the
existence of a real-linear function $\xvar(f)$ such that
\be
\left[b,\hat A(f)\right]_-=\xvar(f)\Io.
\ee
This implies
\be
\sigma^\times(f,0;0,g')&=&\frac{i}{2}\left[\hat A(f),\hat A^{\rm j}(g')\right]_-\cr
&=&\Im\left(\xvar(f)\overline {\yvar(g')}\right).
\ee
The obvious choice of symplectic form $\sigma^\times(f,f';g,g')$
is then
\be
\sigma^\times(f,f';g,g')
&=&\sigma(f,g)+\sigma^{\rm j}(f';g')\cr
& &
+\Im\xvar(f)\overline{\yvar(g')}
+\Im\yvar(f')\overline{\xvar(g)}.
\label {sigmacross}
\ee
The positivity requirement (\ref {posreq}) suggest now to
define
\be
s^\times(f,f';g,g')&=&s(f,g)+s^{\rm j}(f',g')\cr
& &
+\Re\xvar(f)\overline{\yvar(g')}
+\Re\yvar(f')\overline{\xvar(g)}.
\ee
Positivity is satisfied provided
\be
|x(f)|^2\le s(f,f)=\langle\phi|\phi\rangle.
\ee
This implies the existence of functions $f^{(1)}_x$ and $f^{(2)}_x$ satisfying
\be
s(f^{(1)}_x,f^{(1)}_x)+s(f^{(2)}_x,f^{(2)}_x)\le 1,
\ee
for which
\be
x(f)=s(f^{(1)}_x,f)+is(f^{(2)}_x,f).
\ee

The symplectic form $\sigma^\times(f,f';g,g')$ and the bilinear form $s^\times(f,f';g,g')$
together determine the correlation function ${\cal F}^\times(f,f';g,g')$
via (\ref {prodcorfun}), and a corresponding state
of the covariance system with group $G\times G$. The diagonal ${\cal F}^{\rm I}(f,g)$
describes a state of the quantized e.m.~field interacting
with a quantum current. It is the latter state which is analyzed below.

\section{A quantum source of e.m. radiation}

First of all note that the state determined by ${\cal F}^{\rm I}(f,g)$
from a classical point of view describes always a vacuum. Indeed,
from
\be
{\cal F}^{\rm I}(f,0)&=&{\cal F}(f,0)e^{(1/2)|\xvar(f)|^2}e^{-(1/2)|\xvar(f)-\yvar(f)|^2}
\ee
follows by expansion to first order in $f$ that
\be
\langle A^{\rm I}(f)\rangle=0.
\label {vanishingfield}
\ee
The latter quantity is the classical part of the smeared-out e.m.~vector
potential. That it vanishes is in agreement with the pure quantum
nature of the currents $\hat j_\mu(q)$ whose quantum expectation vanishes as well. 

Next consider the scattering situation with a current localized in spacetime,
i.e.~$\alpha_\mu(q)=0$ outside some bounded region in the vicinity of the origin of
spacetime. In addition, let the Green's function $\Delta_G(q)$ in (\ref {yform})
be the retarded Green's function. Then $\yvar(f)=0$ holds for all $f$
with support in the far past. As a consequence, for $f',g'$ with 
support in the far past is ${\cal F}^\times(f,f';g,g')={\cal F}(f,g)$.
Hence, the state of the system in the past is the vacuum of the free e.m.~field.

The function $\yvar(f)$ does not vanish for all $f$ with support in the future,
even when $\alpha_\mu(q)$ is again zero. In other words, the quantum current
produces a radiation field. A first observation is that this radiation field cannot be
coherent because the square of $\yvar(f)$ appears in (\ref {fj}).
This is also obvious from (\ref {vanishingfield}).
Nontrivial quantum fluctuations are present, as can be seen by
expanding ${\cal F}^{\rm I}(f,g)$ to first order in $f$ and $g$
\be
\langle\hat A^{\rm I}(g)\hat A^{\rm I}(f)\rangle
&=&\langle\phi|\psi\rangle-\overline{x(f)}y(g)-\overline{y(f)}x(g)
+\overline{y(f)}y(g).
\ee
The first term in the r.h.s.~describes the vacuum fluctuations. The last term
describes field fluctuations which are identical to those of a classical
radiation field. This leads to the remarkable observation that in this
model quantum currents produce e.m.~fields which propagate like classical
radiation fields. They differ from them because their quantum expectation
vanishes. Whether such quantum radiation fields exist in nature
is not immediately clear. In more sophisticated models one can
expect that these quantum radiation fields will survive in some sense.
Indeed, any quantum current $\hat j_\mu(q)$ can be decomposed
into the sum of a classical current $\langle \hat j_\mu(q)\rangle$ and a
remainder with vanishing average. Now assume that the interacting
field operators $\hat A^{\rm I}_\mu(q)$ satisfy the d'Alembert equation.
Then, by linearity, the field operators decompose into the sum
of a field produced by the classical current and a remainder without
classical analogue, of the type found in the present model.

A limitation of the model is that the quantum current is described by
one single harmonic oscillator. As a consequence, the interacting field operators
$\hat A^{\rm I}(f)$ satisfy unsatisfactory commutation relations (see (\ref {comrel})
and (\ref {sigmacross}))
\be
\left[\hat A^{\rm I}(f),\hat A^{\rm I}(g)\right]_-
&=&-2i\sigma^{\rm I}(f,g)
\ee
with
\be
\sigma^{\rm I}(f,g)&=&\sigma(f,g)-\Im x(f)\overline{x(f)}\cr
& &+\Im\big(x(f)+y(f)\big)\overline {\big(x(g)+y(g)\big)}.
\ee
It is tempting to modify the model in such a way that
the first two contributions to $\sigma^{\rm I}(f,g)$ cancel.
One can hope that in such a modified model the energy density
of the vacuum, which is infinite in absence of interactions,
becomes finite in presence of quantum fields produced by
quantum currents.

\section{Conclusions}

The present paper develops a rigorous theory for quantized e.m.~fields
interacting with given current fields. The formalism of
covariance systems is used. The main tool in this approach is the vacuum to
vacuum correlation function.
It is not assumed {\sl a priori} that the
field operators are those of the non-interacting theory.
Instead they are the sum of the free field operator and of
a field operator which is the solution of the d'Alembert
equation with the given quantum current as source term.

The theory has been applied to two simple models, the first
of which is the well-known model of quantized e.m.~fields
interacting with a classical current. We show that our approach
agrees with standard results. In case the currents vanish
outside a finite part of spacetime then
our result describes radiation fields which correspond with
coherent states, at least, if no infrared divergency occurs.
The second model describes a current whose quantum expectation
vanishes. It produces a radiation field whose quantum expectation
vanishes as well. More sophisticated models are expected to
produce similar results. 

Our main conclusion is that the algebra of field operators depends on details
of the applied current. In this aspect our work goes beyond
other approaches based on fixed algebraic structures
and their representations. The emphasis on correlation functions
solves the related technical problems in an elegant way.


Field operators in the present paper are smeared out with
test functions over spacetime, and {\sl not} with
test functions over 3-dimensional space, as is often
done. We were not able to make a transition between these two
approaches. In particular, we did not obtain a Heisenberg picture
with a Hamiltonian, dependent on the external current,
describing the time evolution of field operators smeared out
with test functions over 3-dimensional space. If it turns out
that such description does not exist, then this is bad news
for the standard approach, based on scattering theory,
which takes the existence of an interaction
picture for granted.


The present work opens perspectives which may eventually lead to
a rigorous formalism of quantum electrodynamics. The next step
to take along the lines of the present paper is a rigorous
description of a field of Dirac electrons interacting with
a classical e.m.~field. The algebra of Dirac currents is
more complicated than what is supported by the present paper.
The resulting technical problems have to be solved as well.

\section*{Acknowledgments}
We thank Prof. H. Spohn for drawing our attention to \cite {KTWB68}.
We thank an unknown referee for pointing out ref.~\cite {PS95}.
We acknowledge fruitful discussions with M. Czachor and M. Kuna
about the topic of this paper.

{
  \appendix
  \renewcommand{\theequation}{A\arabic{equation}}
  \setcounter{equation}{0}  
  \section*{Appendix A}

Here we give a justification for assumption (\ref{wfl}).

Assume two vector potentials differ only by a gauge transformation
\be
A'_\mu(q)=A_\mu(q)+\partial_\mu \chi(q),
\ee
with $\chi(q)$ an arbitrary solution of the d'Alembert equation.
Then the smear\-ed-out fields satisfy
\be
A'(f)-A(f)
&=&\int_{\Ro^4}{\rm d}q\,f^\mu(q)\partial_\mu \chi(q)\cr
&=&i\int_{\Ro^4}{\rm d}k\,\tilde \chi(-k)\tilde f^\mu(k)k_\mu.
\ee
Since the two vector potentials are physically equivalent, it should not
be possible to distinguish them by means of the test function $f$.
The condition $A'(f)=A(f)$ implies then that (\ref{wfl}) must hold
for all wave vectors $k$ satisfying $k^\mu k_\mu=0$.

Assume now that $\tilde g_\mu(k)$ satisfies $k^\mu g_\mu(k)$
whenever $k^\nu k_\nu=0$, in such a way that the function
\be
\chi(k)=\frac{1}{k^\nu k_\nu}k^\mu g_\mu(k).
\label{chidef}
\ee
remains continuous. Then one can decompose $\tilde g_\mu(k)$
into parts parallel and orthogonal to the wave vector
\be
\tilde g_\mu(k)=\tilde f_\mu(k)+k_\mu \tilde \chi(k)
\ee
where $\tilde f_\mu(k)$ satisfies (\ref{wfl}) for all $k$.
One verifies immediately that $A(g)=A(f)$ holds for any
vector potential $A$. Hence, we can always chose the test functions
$f$, satisfying (\ref{wfl}) for all $k$, as representative
for a whole class of equivalent test functions $g$, satisfying (\ref{wfl})
when $k^\mu k_\mu=0$, and such that (\ref {chidef}) remains continuous.

}
{
  \appendix
  \renewcommand{\theequation}{B\arabic{equation}}
  \setcounter{equation}{0}  
  \section*{Appendix B}

Here we show that the correlation function (\ref{corfun})
satisfies the necessary conditions.
The extension of the arguments to correlation function (\ref{intcorfun})
is straightforward.

Normalization ${\cal F}(0,0)=1$ is clear. Positivity follows from
\be
\sum_{mn}\overline{\lambda_n}\lambda_n{\cal F}(f_m,f_n)
&=&\sum_{mn}\overline{\lambda_n}\lambda_n
e^{\langle f_m|\,f_n\rangle}
e^{-(1/2)\langle f_m|\, f_m\rangle}
e^{-(1/2)\langle f_n|\,f_n\rangle}\cr
&=&\sum_{mn}\overline{\mu_n}\mu_n
e^{\langle f_m|\,f_n\rangle}
\label{appa1}
\ee
with
\be
\mu_m=\lambda_m e^{-(1/2)\langle f_m|\, f_m\rangle}.
\ee
Note that the matrix with elements $\langle f_m|\,f_n\rangle$
is positive definite.
Hence, positivity of (\ref{appa1}) follows by means of Schur's lemma.

Finally, covariance follows from
\be
{\cal F}(f+h,g+h)&=&e^{i\Im\langle f+h|\,g+h\rangle}
e^{-(1/2)\langle f-g|\, f-g\rangle}\cr
&=&e^{i\Im\langle h|\,g\rangle}e^{i\Im\langle f|\,h\rangle}
{\cal F}(f,g).
\ee

}

\end{document}